\documentclass{PoS}
\usepackage{graphicx,amsmath,amssymb,dcolumn,bm}

\title{Standard convolution description of deuteron tensor spin structure}

\ShortTitle{Standard convolution description of deuteron tensor spin structure}

\author{\speaker{W. Cosyn}\\
        Department of Physics and Astronomy, Ghent University, Proeftuinstraat 
86, B9000 Ghent, Belgium\\
        E-mail: \email{wim.cosyn@ugent.be}}

\author{Yu-Bing Dong\\
       Institute of High Energy Physics, Chinese Academy of Sciences, Beijing 
100049, China\\
Theoretical Physics Center for Science Facilities (TPCSF), CAS, Beijing 100049, 
China\\
  }

\author{S. Kumano\\
       KEK Theory Center, Institute of Particle and Nuclear Studies,
High Energy Accelerator Research Organization (KEK),
1-1, Ooho, Tsukuba, Ibaraki, 305-0801, Japan\\
J-PARC Branch, KEK Theory Center, Institute of Particle and Nuclear Studies, 
KEK,
and Theory Group, Particle and Nuclear Physics Division, J-PARC Center,
203-1, Shirakata, Tokai, Ibaraki, 319-1106, Japan
}

\author{M. Sargsian\\
       Department of Physics, Florida International University, Miami, Florida 
33199, USA
}       
       
\abstract{Spin-1 hadrons have additional structure functions not present for 
spin 1/2 hadrons.  These could probe novel aspects of hadron structure and QCD 
dynamics.  For the deuteron, the tensor structure function $b_1$ 
inherently mixes quark and nuclear degrees of freedom.  These proceedings 
discuss two standard convolution models applied to calculations of the deuteron 
$b_1$ structure functions.  We find large differences with the existing HERMES 
data and other convolution model calculations.  This leaves room for 
non-standard contributions to $b_1$ in the deuteron.  We also discuss the 
influence of higher twist nuclear effects in the model calculations and data 
extraction at kinematics covered in HERMES and Jefferson Lab.}

\FullConference{XXV International Workshop on Deep-Inelastic Scattering and Related Subjects\\
		3-7 April 2017\\
		University of Birmingham, UK}

\begin{document}

\section{Introduction}
In addition to vector spin observables familiar from the spin 1/2 case,  a spin 
1 hadron also gives access to additional tensor spin observables.  In inclusive 
deep inelastic scattering (DIS) these give rise to four additional structure 
functions, called $b_{1-4}$~\cite{Hoodbhoy:1988am}.  Two ($b_1,b_2$) of these 
are leading twist and obey a Callan-Gross like relation $b_2=2x_Tb_1$, where 
$x_T=Q^2/2Pq$ is the Bjorken scaling variable for the spin 1 hadron.  In the 
parton model, $b_1$ obeys a sum rule $\int dx b_1(x)=0$~\cite{Close:1990zw} 
when considering only the valence quark sector and $b_1$ has an explicit 
interpretation as a function of unpolarized quark distributions in a polarized 
hadron
\begin{equation}
 b_1=\frac{1}{2}\sum_q e_q^2(q^0-q^1)\,,
\end{equation}
where the sum runs over all (anti)quark flavors, $e_q$ is the fractional quark 
charge and $q^i$ represents the unpolarized quark distribution function in a 
hadron with 
polarization $i$.

Experimentally, $b_1$ can be extracted in polarized inclusive DIS from 
measuring the tensor asymmetry
\begin{equation}
 A_{zz}=\frac{\sigma^++\sigma^--2\sigma^0}{\sigma^++\sigma^-+\sigma^0}\,,
\end{equation}
where $\sigma^i$ is the cross section for a target with polarization $i$ along 
a chosen direction.  For the deuteron, Hermes measured 
$A_{zz}$~\cite{Airapetian:2005cb} and found a sizeable asymmetry and hence 
also extracted $b_1$ in its covered kinematics. In the near future, the 12 GeV 
upgrade of 
Jefferson Lab will probe tensor polarization in the deuteron in two 
experiments~\cite{Slifer:2013vma}, one in the DIS regime that will improve 
experimental knowledge of $A_{zz}$ and $b_1$, the second in the quasi-elastic 
regime.  Additionally, opportunities to access tensor polarization in the 
deuteron exist at Fermilab in Drell-Yan reactions~\cite{Kumano:2016ude} and 
at the Jefferson Lab implementation of a future electron ion collider (JLEIC), 
also allowing for spectator nucleon tagging 
capabilities ~\cite{Cosyn:2014zfa,Cosyn:2016oiq}. 

In standard calculations of the deuteron, 
considering only the $pn$-component, $b_1$ is only non-zero because of the 
$D$-wave component in the nuclear wave function.  Due to the small size of the 
$D$-wave component, the obtained $b_1$ is very small and cannot explain the 
size of the HERMES data.  This suggests the need to consider more advanced or 
exotic mechanisms, such as 
shadowing~\cite{Frankfurt:1983qs,Nikolaev:1996jy,Edelmann:1997ik,Bora:1997pi}, 
eikonal final-state interactions~\cite{Cosyn:2017ekf}, and pionic and hidden 
color contributions~\cite{Miller:2013hla}, where inclusion of the latter can 
explain the HERMES data.  Model calculations for $b_1$ considering the 
$pn$ component~\cite{Hoodbhoy:1988am,Khan:1991qk} are scarce in the literature 
but are essential to constrain the baseline calculation. 
 Recently, we calculated $b_1$ in two standard convolution 
models~\cite{Cosyn:2017fbo}, and found significant deviation from the previous 
model calculations.  The formalism and results of these calculations are 
summarized in the following sections, for more details we refer to 
Ref.~\cite{Cosyn:2017fbo}.

\section{Standard convolution formalism for $b_1$ in two approaches}

For nuclear DIS in a standard convolution formulation, separation of scales 
between nuclear and partonic structure is used to write the nuclear hadronic 
tensor $W^A_{\mu\nu}$ as a convolution of a nuclear spectral function $S(p)$ 
and the 
hadronic tensor of the nucleon $W^N_{\mu\nu}$:
\begin{equation}
 W^A_{\mu\nu}(P_A,q)=\int d^4pS(p)W^N_{\mu\nu}(p,q)\,.
\end{equation}
In a first approach (Theory 1), scaling limit relations between virtual 
photon-hadron helicity amplitudes and structure functions of the deuteron and 
nucleon
are used to obtain the expression
\begin{equation}
 b_1(x,Q^2)=\int 
\frac{dy}{y}\left[f^0(y)-\frac{f^+(y)+f^-(y)}{2}\right]F^N_1(x/y,Q^2)\,,
\label{eq:b1conv}
\end{equation}
where $F_1^N=(F^p_1+F^n_1)/2$ is the average of proton and neutron structure 
functions, and
\begin{equation}
 f^H(y) = \int 
d^3\bm p\,y\,|\phi^H(\bm p)|^2\delta\left(y-\frac{\sqrt{m_N^2+\bm 
p^2}-p_z}{m_N} \right)\,,
\end{equation}
with $\phi^H(\bm p)$ the deuteron wave function for polarization $H$, 
normalized as $\int d^3\bm p\,y\,|\phi^H(\bm p)|^2=1$.  For the nucleon 
$F_1^N$, the leading order expression, taking into account the finite ratio of 
transverse to longitudinal cross sections $R=\sigma_L/\sigma_T$ is used
\begin{equation}
 F^N_1(x,Q^2)=\frac{1+4m_N^2x^2/Q^2}{2x[1+R(x,Q^2]}\, x\sum_f e^2_f \left[ 
q_f(x,Q^2)+\bar{q}_f(x,Q^2)\right]_{\text{LO}}\,.
\end{equation}

A second approach (Theory 2) is based on the virtual nucleon approximation 
(VNA) framework, which has been applied previously to unpolarized deuteron 
DIS~\cite{Cosyn:2010ux,Cosyn:2013uoa} and can be generalized to polarized 
reactions.  In the VNA approach, no scaling limit relations are assumed, hence 
higher twist nuclear effects are automatically included.  The VNA expression 
for $b_1$ is given by
\begin{multline}\label{eq:b1vna}
 b_1(x,Q^2)=\frac{3}{4(1+Q^2/\nu^2)} \int \frac{k^2}{\alpha_i} 
dk \, d(\cos\theta_k)
\left[ F_{1}^N(x_i,Q^2) \left(6\cos^2\theta_k-2\right)
\right.\\ \left. 
+\frac{\bm p_i^{\perp 2} } {2 \, p_i q } F_{2 }^N (x_i ,Q^2)
\left(5\cos^2\theta_k-1\right) \right]
\left[ \frac{U(k)W(k)}{\sqrt{2}}+\frac{W( k)^2}{4}\right] .
\end{multline}
Here $\nu$ is the virtual photon energy in the deuteron rest frame, $p_i$, 
$x_i=Q^2/2p_iq$ and $\alpha_i=2p_i^-/P^-$ are respectively the four-momentum, 
Bjorken 
variable and lightcone momentum fraction of the struck nucleon, $k$ is the 
dynamical variable appearing in the light-front deuteron wave 
function related to the deuteron and nucleon momenta by~\cite{Frankfurt:1981mk}
\begin{align}
 &k^3=(1-\alpha_i)E_k &E^2_k=\frac{m_N^2+\left(\bm 
p_i^\perp+\frac{\alpha_i}{2}\bm P^\perp\right)}{\alpha_i(2-\alpha_i)}\,.
\end{align}
$U(k)$, $W(k)$ are the radial $S$- and $D$-wave components of the 
light-front deuteron wave function obeying the baryon and momentum sum rules
\begin{align}
 &\int \frac{d\bm k}{E_k}\left[U(k)^2+W(k)^2\right]=1 &\int \frac{d\bm 
k}{E_k}\,\alpha_i\,\left[U(k)^2+W(k)^2\right]=1\,,
\end{align}
and are here approximated by their non-relativistic 
counterparts.  Comparing Eq.~(\ref{eq:b1vna}) with Eq.~(\ref{eq:b1conv}), the 
presence of the 
additional 
$F_2^N$ term reflects the inclusion of higher twist nuclear effects.

\begin{figure}[ht]
\begin{center}
\includegraphics[width=.45\textwidth]{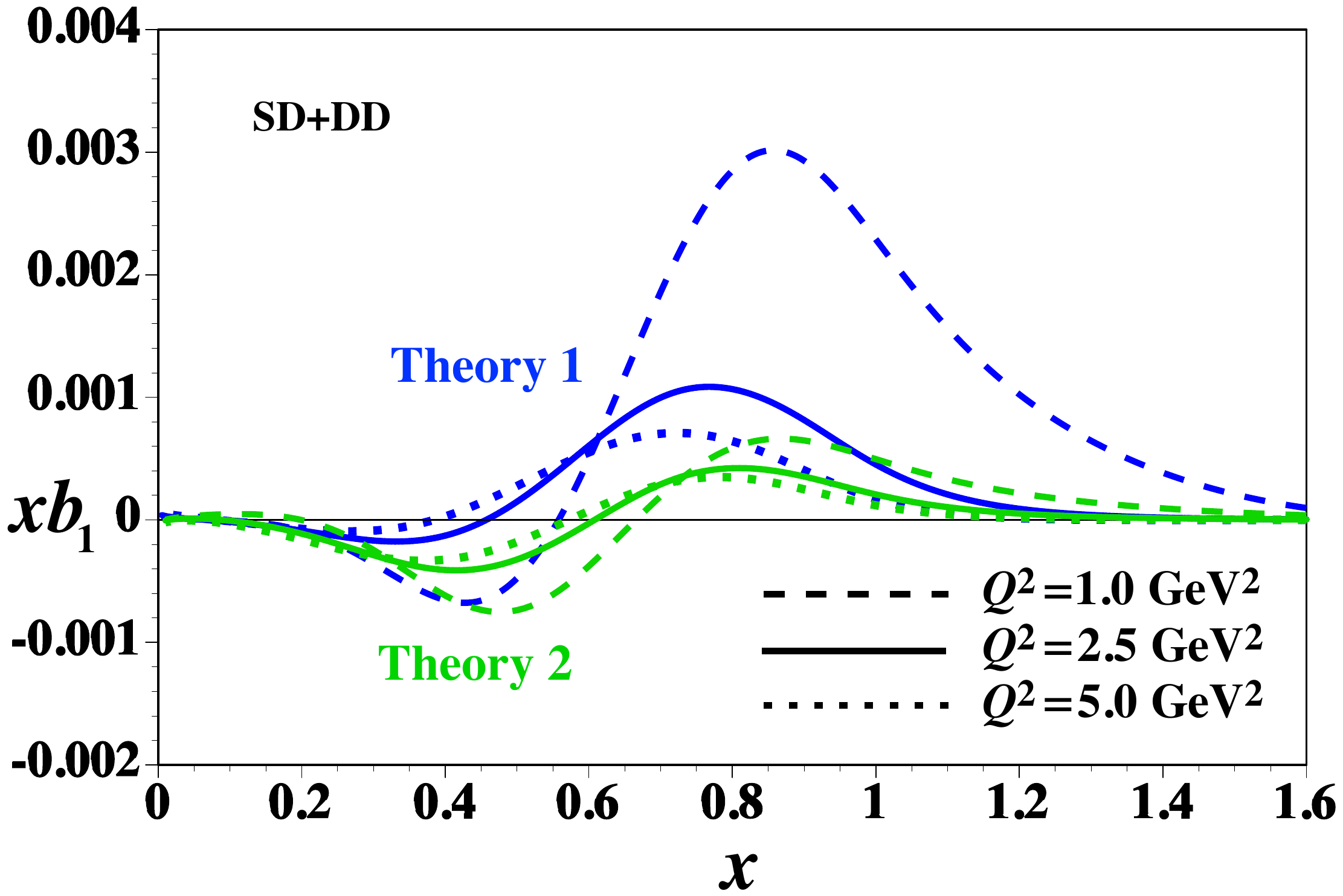}
\qquad
\includegraphics[width=.45\textwidth]{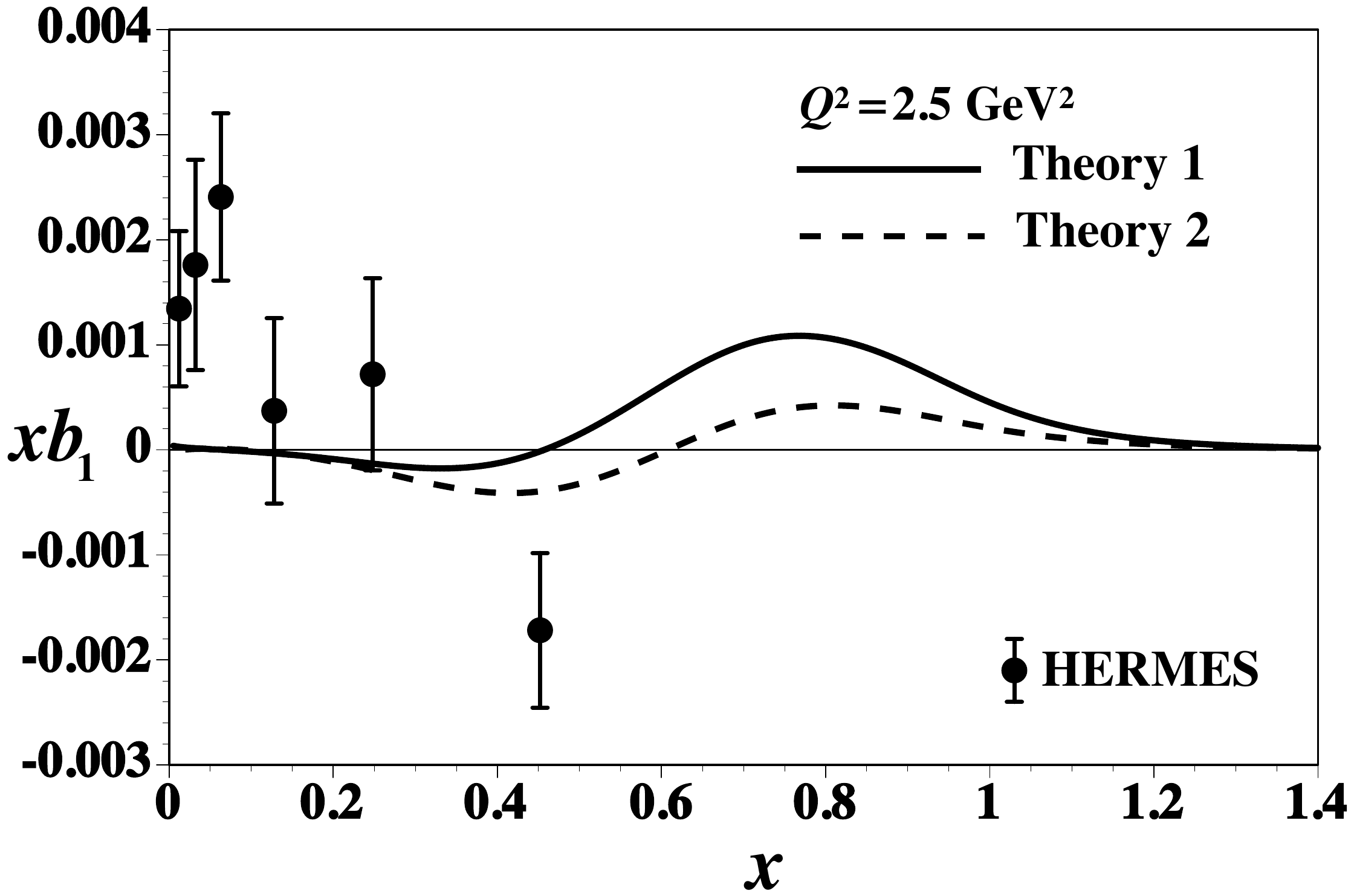}
\end{center}
\caption{Calculations of deuteron structure function $b_1$ by the two
convolution descriptions of Theory 1 [Eq.~(\ref{eq:b1conv})] and Theory 2 
[Eq.~(\ref{eq:b1vna})].  
(Left panel) $Q^2$ dependence of $xb_1$ at $Q_2$ 1.0, 2.5,
and 5.0 GeV$^2$. (Right panel)  Comparison with the HERMES 
data~\cite{Airapetian:2005cb}.  Calculations are for $Q^2=2.5~\text{GeV}^2$, 
representative for the average $Q^2$ value of the HERMES data.  Figure adapted 
from Ref.~\cite{Cosyn:2017fbo}}
\label{fig:b1_q2}
\end{figure}

\section{Results}

Fig.~\ref{fig:b1_q2} shows the $Q^2$-dependence of the deuteron $b_1$ for 
the two different calculations and compares the calculations to the HERMES data. 
 In these calculations, we used the MSTW2008 (Martin-Stirling-Thorne-Watt,
2008) leading-order (LO) parametrization for $F_N^2$, the SLAC-
R1998 parametrization for the ratio $R$, and the CD-Bonn deuteron wave 
function. We observe that both calculations exhibit a 
similar oscillating $x$-dependence.  Compared to the calculations of 
Ref.~\cite{Hoodbhoy:1988am,Khan:1991qk} (denoted KH from now on) two differences 
are worth noting: (i) the dominant term originating from the deuteron $SD$-wave 
interference (not shown separately here, see Fig.~4 of 
Ref.~\cite{Cosyn:2017fbo} ) has an opposite sign in our calculations compared to 
the KH 
calculations, (ii) we find a non-zero $b_1$ for $x>1$, whereas it 
is identically zero in the KH results.  The left panel of Fig.~\ref{fig:b1_q2} 
shows that the difference in size between the two calculations becomes larger 
for smaller $Q^2$ values.  The main origin of this is the inclusion of higher 
twist effects in Theory 2.  Another origin is the different way deuteron 
nuclear structure is considered (wave function normalization, instant form 
versus light-front form wave function).  The variation of the deuteron $b_1$ 
with $Q^2$ shows its sensitivity to dynamical aspects of hadron structure. 
When comparing our calculations with the HERMES data in the right panel of 
Fig.~\ref{fig:b1_q2}, we see that both calculations fail to accurately describe 
the data, 
though it has to be noted the error bars are quite large.  The upcoming 
Jefferson Lab data should improve that in the future.  Nevertheless, this 
current comparison certainly does not rule out the possibility of additional 
mechanisms (possibly of exotic nature) playing an important role in the $b_1$ 
of the deuteron.

Another point worth of scrutiny is the way $b_1$ is extracted from the $A_{zz}$ 
observable.  For the HERMES experiment, this was done using formulas that 
include Bjorken scaling limit relations and neglect the higher twist $b_3,b_4$. 
 Our analysis (see Ref.~\cite{Cosyn:2017fbo}) shows that this is not 
necessarily the case for the kinematics of HERMES and Jefferson Lab, with the 
Callan-Gross like relation violated and the higher twist $b_{3,4}$ of similar 
magnitude as the leading twist structure functions. Consequently, 
inclusion of higher twist effects in the extraction procedure might be 
warranted at these kinematics to accurately extract $b_1$.

\section{Conclusion}
We have summarized calculations of the $b_1$ deuteron structure function in 
two models based on the standard convolution approach of nuclear DIS.  We find 
significant differences with older calculations and our calculations cannot 
reproduce the size or trend of the HERMES data, leaving room for more advanced 
or exotic
mechanisms playing an important role.  An upcoming experiment at Jefferson Lab 
and additional opportunities at Fermilab and a future JLEIC could shed more 
light on these issues and motivate additional theoretical work.

\acknowledgments

This work was supported by Japan Society for the Promotion of Science (JSPS)
Grants-in-Aid for Scientific Research (KAKENHI) Grant No. JP25105010. It is 
also partly supported by the National Natural Science Foundation of China (No. 
11475192) and by the fund of the Sino-German CRC 110 ``Symmetries and the 
Emergence of Structure in QCD project'' (NSFC Grant No. 11621131001). Y.-B. D.  
thanks the warm hospitality of the KEK theory center during his visit.

\bibliography{../bibtexall.bib}

\end{document}